\documentclass[conference]{IEEEtran}

  \usepackage[pdftex]{graphicx}
  \DeclareGraphicsExtensions{.pdf}

\usepackage{algorithmic}
\usepackage{algorithm}
\usepackage{amssymb}

\newtheorem{Theorem}{Theorem}

\newtheorem{Proposition}{Proposition}

\newtheorem{Remark}{Remark}

\begin{document}

\IEEEoverridecommandlockouts
\title{Maximum Multiflow in Wireless Network Coding
\thanks{This work is supported by the NSFC projects (60972011, 61003100), the 973
project of of China (2012CB315803), and the Research Fund for
the Doctoral Program of Higher Education of China (20100002110033, 20100002120018).}}
\author{\IEEEauthorblockN{Jin-Yi Zhou \quad Shu-Tao Xia\quad Yong Jiang \quad Hai-Tao Zheng}\IEEEauthorblockA{Graduate School at Shenzhen, Tsinghua University, Shenzhen 518055, China\\
\{zhoujy09@mails, xiast@, jiangy@, zheng.haitao@\}sz.tsinghua.edu.cn}}

\maketitle

\begin{abstract}
In a multihop wireless network, wireless interference is crucial to the maximum multiflow (MMF) problem, which studies the maximum throughput between multiple pairs of sources and sinks. In this paper, we observe that network coding could help to decrease the impacts of wireless interference, and propose a framework to study the MMF problem for multihop wireless networks with network coding. Firstly, a network model is set up to describe the new conflict relations modified by network coding. Then, we formulate a linear programming problem to compute the maximum throughput and show its superiority over one in networks without coding. Finally, the MMF problem in wireless network coding is shown to be NP-hard and a polynomial approximation algorithm is proposed.
\end{abstract}

\IEEEpeerreviewmaketitle

\section{Introduction}

In recent years, there has been a focus on the maximum multiflow (MMF) problem
in multihop wireless networks \cite{impact}\cite{mmf}, which studies the maximum throughput between multiple pairs of sources and sinks. Wireless interference is a key issue affecting the performance of multihop wireless networks.  Jain {\it et al}. \cite{impact} introduce a {\it conflict graph} to model the effects of wireless interference, and compute the maximum throughput by formulating the MMF problem under the constrains derived from the conflict graph. Wan \cite{mmf} shows that solving the MMF problem is NP-hard and provides polynomial approximation algorithms on MMF problems in multihop wireless networks.

It is well known that network coding \cite{nif} is an effective technique to improve the throughput of networks. COPE \cite{cope}, as a new architecture of multihop wireless network, demonstrates the practical throughput gain by employing network coding. Consequently, there has also been a focus on computing the maximum throughput or throughput gain of a wireless network with network coding \cite{bound1}-\cite{bound4}. However, these previous works assume many conditions which are sometimes unpractical, such as homogeneity of node, randomness of network topology, regularity of communication pattern, or optimal scheduling. For any given multiple flows in any multihop wireless network with network coding, formulating the MMF problem and computing the maximum throughput are challenging problems.

By performing network coding, a wireless node could merge multiple transmissions that may interfere with each other into a single one. That is to say, network coding could help to decrease the impacts of the wireless interference in multihop wireless networks. Based on the idea, in this paper, we introduce a new conflict graph to model the wireless interference under network coding, and rebuild the constraints derived from the new conflict graph when formulating the MMF problem in multihop wireless networks with network coding. Moreover, we show that the MMF problem in a multihop wireless network with network coding is also NP-hard, and propose a polynomial approximation algorithm.

{\it Major Contributions}: 1) Introduce a general network model for multihop wireless networks with or without network coding; 2) Propose a framework to compute the maximum throughput for any given multiple unicast flows in any multihop wireless network with network coding; 3) Propose a polynomial approximation algorithm in multihop wireless networks with network coding.

The paper is organized as follows. Preliminaries are given in Section II. We illustrate the impacts of network coding on wireless interference and introduce a new network model in Section III. The formulation of the MMF problem in wireless network coding is in Section IV and the polynomial approximation algorithm is proposed in Section V. Finally, we make some discussions and conclusions in Section VI.

\section{Preliminaries}

In this section, without considering network coding, we describe the network model for multihop wireless networks, and explain the corresponding MMF problem and the capacity region supported by these networks \cite{impact}\cite{mmf}\cite{mcmr}.

\subsection{Network Model}
For simplicity, the multihop wireless network discussed in this paper is assumed to be a {\it single-channel single-radio} multihop wireless network under the {\it protocol model} of interference \cite{wnc}, in which all antennas are {\it omnidirectional} and all channels have the same bandwidth normalized to one. Some generalizations, e.g., arbitrary link bandwidth, will be discussed in Section VI. Let $S$ be a finite set, $|S|$ denote its cardinality, and ${\mathbb R}_+$ be the set of non-negative real numbers. A function $f$ from $S$ to ${\mathbb R}_+$ is denoted by $f \in {\mathbb R}_+^S$ for short. For any subset $S' \subseteq S$, $f(S') \triangleq \sum_{e \in S'} f(e)$.

Let $(N,A,G,{\cal I})$ represent a multihop wireless network, where $N$ is the set of nodes, $A$ the set of links, $G$ the conflict graph, $\mathcal{I}$ the collection of schedulable link sets or independent sets. The explanation follows in details.
An instance of multihop wireless networks is specified by a finite set $N$ of nodes together with a communication radius function $r \in \mathbb R_+^N$ and an interference radius function $\rho \in \mathbb R_+^N$, where $\rho(i) \ge r(i)$ for any $i\in N$. Let $d_{ij}$ be the Euclidean distance between nodes $i,j$. There is a (communication) link $(i,j)$ from $i$ to $j$ if and only if $0 < d_{ij} \le r(i)$. Let $A$ be the set consisting of all links $(i,j)$ for $i,j \in N$. Then, the communication topology of the multihop wireless network can be represented by the directed graph $(N,A)$. Two links $(i,j),(i',j') \in A$ are said to be {\it conflicted} if $d_{i'j} \le \rho(i')$ or $d_{ij'} \le \rho(i)$. In order to describe the conflict relations of links of $A$, the so-called \emph{conflict graph} $G=(V,E)$ is defined, where $V$ is a finite set of vertexes each of which denotes a link of $A$ and vice versa, and there is an edge $e\in E$ between vertexes $v,v'\in V$ if and only if the links corresponding to $v,v'$ are conflicted. Clearly, the conflict graph $G$ is undirected and $|V|=|A|$. A set of vertexes $V' \subseteq V$ is said to be {\it independent} in $G$ if there is no edge between any pair of vertexes in $V'$. The set of links $L \subseteq A$ is called a {\it schedulable link set} if the corresponding vertex set $V_L\subseteq V$ is independent. This is because the links in $L$ is \emph{conflict-free} and could be scheduled simultaneously when $V_L$ is independent in $G$. Let $\cal I$ be the collection of all independent sets in $G$, i.e., $\cal I$ consists of all schedulable link sets of $A$.

For a multihop wireless network $(N,A,G,\cal I)$, arrange the links of $A$ by a fixed order and index it by $1,2,\ldots, |A|$. For a link set $S\subseteq A$, its incidence vector is a binary vector with length $|A|$ and the $i$-th component is 1 if the $i$-th link is in $S$ and 0 otherwise for $i=1,2,\ldots,|A|$. Let $P$ be the convex hull of the incidence vectors of all independent sets in $\cal I$, called the \emph{independence polytope} of $(N,A,G,\cal I)$.

A \emph{fractional link schedule}, denoted by $\cal S$, is defined by a sequence of pairs $\{(I_j,\lambda_j) \in {\cal I}\times {\mathbb R}_+:1 \le j \le k\}$, where $k$ is a positive integer and $\sum _{j=1}^k \lambda_j \le 1$. The value $\sum _{j=1}^k \lambda_j$ is called the \emph{length} of $\cal S$. The \emph{link capacity function} $c_{\cal S} \in \mathbb R_+^A$ of $\cal S$ is defined by
\begin{eqnarray}
c_{\cal S}(a)=\sum_{j=1}^k \lambda_j|I_j \cap \{a\}|=\sum_{j: \,a\in I_j} \lambda_j,\quad\forall a \in A .
\end{eqnarray}
A \emph{link demand function} $d \in {\mathbb R}_+^A$  is said to be \emph{achievable} if there exists a fractional link schedule $\cal S$ such that $c_{\cal S}(a)=d(a)$ for any $a\in A$, and $\cal S$ is called a fractional link schedule of $d$. For an achievable link demand function $d$, the fractional link schedule of $d$ with minimum length is called an {\it optimal fractional link schedule} of $(G,d)$. Clearly, the link demand function $d$ could be regarded as a vector $(d(a),a\in A)$ with length $|A|$. The {\it schedulable polytope} of $(N,A,G,\cal I)$ is defined by the set of achievable link demand functions.

The following result is derived from \cite{impact}.

\begin{Proposition}\label{pro1}\cite{impact} For a multihop wireless network $(N,A,G,\cal I)$, its schedulable polytope is equivalent to its independence polytope $P$.\end{Proposition}

\subsection{Maximum Multiflow and Capacity Region}
For a multihop wireless network $(N,A,G,\cal I)$, let $\delta ^{in}(n)$ and $\delta ^{out}(n)$ denote the set of links of $A$ entering and leaving a node $n \in N$, respectively. Consider two distinct nodes $s,t \in N$. A function $f \in \mathbb R_+^A$ is called a {\it flow} from $s$ to $t$ ({$s$-$t$ flow) if $f\big(\delta ^{in}(n)\big)=f\big(\delta ^{out}(n)\big)$ for any $n\in N\setminus \{s,t\}$. The value of the $s$-$t$ flow $f$ is $val(f)\triangleq f(\delta ^{out}(s))-f(\delta ^{in}(s))$. Suppose that there are $k$ given commodities with pairs $\{s_i,t_i\}$, where $s_i,t_i\in N$ are source and sink for commodity $i$ respectively. Let $\mathcal{F}_i$ denote the set of $s_i$-$t_i$ flows. A $k$-flow is a sequence of flows $<f_1,f_2,\cdots ,f_k>$ with $f_i \in {\cal F}_i,\;i=1,2,\ldots,k.$

The schedulable polytope $P$, consisting of all achievable link demand functions that can be achieved by some optimal fractional link schedules, is also called the \emph{capacity region} of $(N,A,G,\cal I)$. A $k$-flow $<f_1,f_2,\cdots ,f_k>$ is said to be {\it schedulable} if $\sum _{i=1}^k f_i \in P$. Therefore, the MMF problem can be formulated by the following linear program problem
\begin{eqnarray*}
\mbox{(LP-I)}&&\max\;  \sum_{i=1}^k val(f_i), \\
\mbox{s.t.} && f_i \in \mathcal{F}_i, \;i=1,2,\ldots,k, \quad \sum _{i=1}^k f_i \in P.
\end{eqnarray*}

\begin{figure}[t]
\centering{
\includegraphics[width=2.8in]{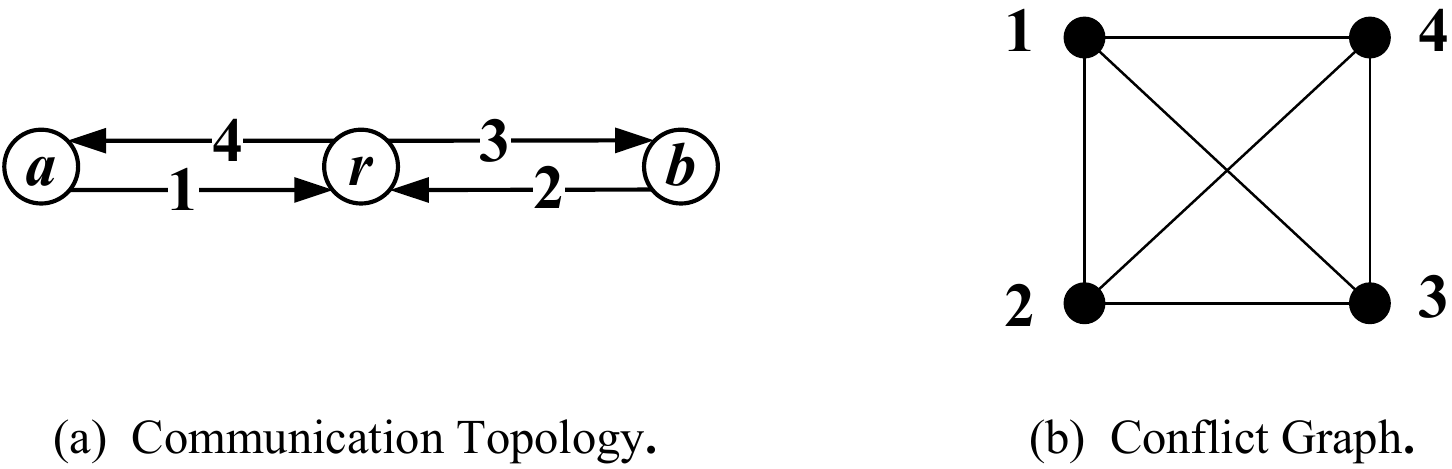}
}
\caption{An example with two commodities $\{a,b\}$ and $\{b,a\}$ which are supported by a multihop wireless network $(N,A,G,\cal I)$, where $N=\{a,r,b\}$, $A=\{1,2,3,4\}$ and ${\cal I}=\{\{1\}, \{2\}, \{3\}, \{4\}\}$. The capacity region $P$ is the convex hull of the following 4 incidence vectors (1000)(0100)(0010)(0001). By formulating the MMF problem, the resulting maximum throughput is ${1}/{2}$ with $val(f_1)=1/4$ and $val(f_2)=1/4$. An optimal fractional link schedule to achieve the maximum throughput is ${\cal S}=\{(1,\frac{1}{4}),(2,\frac{1}{4}),(3,\frac{1}{4}),(4,\frac{1}{4})\}$.}
\label{fig1}
\end{figure}

The maximum throughput is achieved by the MMF problem (LP-I). Fig.\ref{fig1} illustrates an example of computing the maximum throughput for two unicast flows supported by a multihop wireless network.

However, the problem of determining the whole capacity region $P$ of $(N,A,G,\cal I)$, or finding all independent sets of the conflict graph $G$, is NP-hard \cite{mmf}. Moreover, even if the the optimal solution of (LP-I) is obtained, finding an optimal fractional link schedule to achieve it is also NP-hard, which is stated as follows.
Let $\cal N$ be the class of multihop wireless networks discussed in this paper.

\begin{Proposition}\label{pro2} \cite{mmf} Even restricted to the subclass of $\cal N$, in which all nodes have uniform (and fixed) communication radii and interference radii and the positions of all nodes are available, finding an optimal fractional link schedule to meet a given achievable link demand function is NP-hard.\end{Proposition}

Furthermore, Wan \cite{mmf} provided polynomial approximation algorithms on the MMF problem (LP-I).

\section{Network Model with Network Coding}
In this section, we illustrate the impacts of network coding on wireless interference, and introduce a general model for multihop wireless networks with or without network coding.
\subsection{Network Coding}
Network coding could help to overcome the wireless interference in some cases in multihop wireless network. Fig.\ref{fig2} illustrates typical coding scenarios. In Fig.\ref{fig2}a, two packets are exchanged via a relay node. Without network coding, four transmissions are required since they are interferential pairwise. However, after receiving $P_1$ and $P_2$, the relay node can encode or XOR the two received packets and broadcast the encoded packet or $P_1\oplus P_2$. In this way, only three transmissions are required while the interference between the last two transmissions are overcome by the relay node. In Fig.\ref{fig2}b, two packets are crossing a relay node. Without network coding, four transmissions are required since they are also interferential pairwise. When the opportunistic listening is applied, the relay node can encode or XOR the received packets and then broadcast the encoded packet or $P_1\oplus P_2$, and only three transmissions are required. Similarly, the interference between the last two transmissions are overcome by the relay node. Fig.\ref{fig2}c illustrates a more general scenario, where $n$ packets are crossing a relay node. The relay node can encode at most $n$ packets and broadcast the encoded packet in a single transmission, and overcome the interference between any pair of those transmissions. The coding conditions and the upper bound of coding number are given in \cite{cope} and \cite{howmany} respectively. Therefore, by performing network coding, a node can equivalently overcome the interference of involved transmissions, and thus improve the throughput of multihop wireless networks.

Network coding for a multihop wireless network involves two parts. One is the {\it coding nodes} that perform network coding, and the other is the coding coefficients (or \emph{local kernels}) which give detailed coding operations for each coding node. In this paper, the {\it coding structure} of a multihop wireless network with network coding, which refers to coding nodes together with their local kernels, is assumed to be given.

\begin{figure}[t]
\centering{
\includegraphics[width=3.4in]{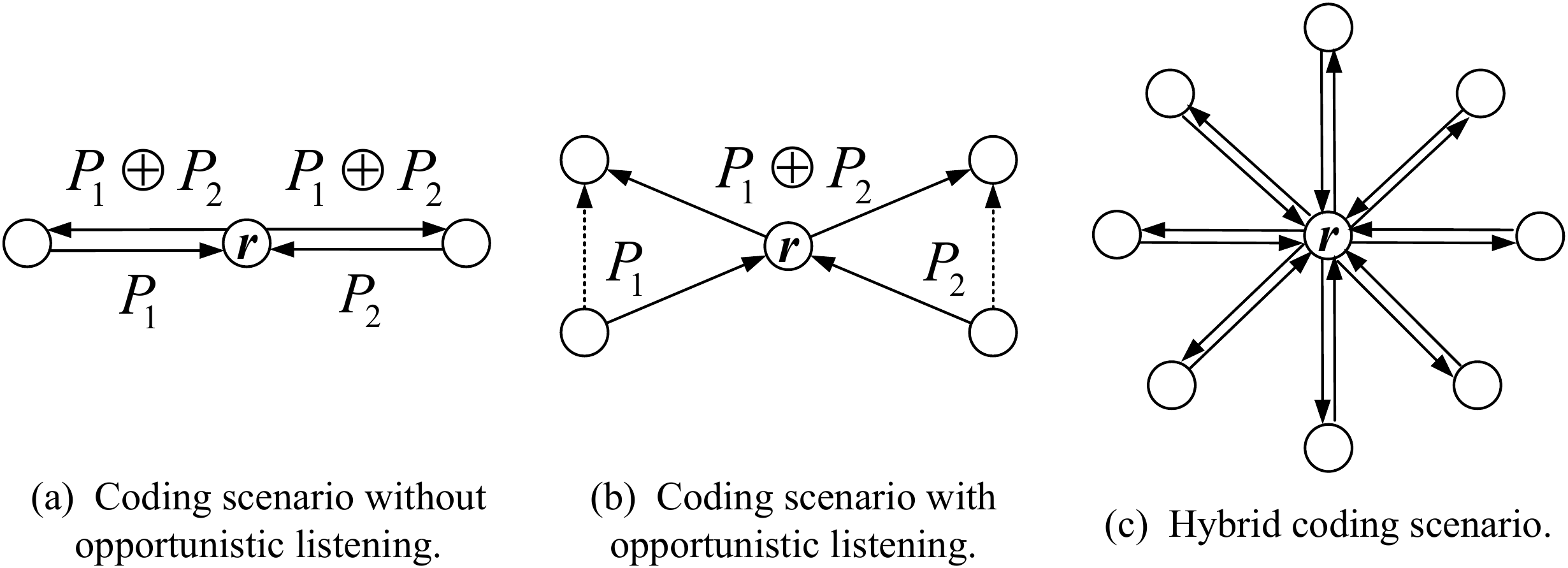}
}
\caption{Typical coding scenarios of wireless network coding}
\label{fig2}
\end{figure}

\subsection{Network Model with Network Coding}
At first, we introduce the concept of {\it hyperarc} to describe the interference relations in multihop wireless networks with network coding.
For a multihop wireless network $(N,A,G,{\cal I})$, by employing network coding, its communication topology $(N,A)$ is actually enhanced by those encoded transmissions. Let $(i,J)$ be a hyperarc for $i \in N$ and $J \subseteq N\setminus i$ satisfying $(i,j)\in A$ for each $j\in J$. Then, each encoded transmission for node $i \in N$ can be represented by a hyperarc $(i,J)$. Let $\hat A$ be the set of hyperarcs representing all possible encoded transmissions in the multihop wireless network with network coding. Note that a hyperarc $(i,J)$ with $|J|=1$ indicates an original (communication) link or an uncoded transmission. For a hyperarc $(i,J) \in \hat A$, a link $(i,j)$ with $j \in J$ is called a {\it sub-link} of $(i,J)$, denoted by $(i,j) \triangleright (i,J)$. Under the protocol model of interference, two hyperarcs $(i,J),(i',J') \in \hat A$ are said to be {\it conflicted} if and only if there is a conflict between any pair of their sub-links $(i,j) \triangleright (i,J)$ and $(i',j') \triangleright (i',J')$.
In order to describe the conflict relations of hyperarcs of $\hat A$, a new conflict graph $\hat G=(\hat V,\hat E)$ is defined, where $\hat V$ is a finite set of vertexes each of which denotes a hyperarc of $\hat A$ and vice versa, and there is an edge $\hat e \in \hat E$ between vertexes $\hat v,\hat v' \in \hat V$ if and only if the hyperarcs corresponding to $\hat v,\hat v'$ are conflicted. Clearly, the conflict graph $\hat G$ is undirected and $|\hat V|=|\hat A|$. A set of vertexes $\hat V' \subseteq \hat V$ is said to be independent in $\hat G$ if there is no edge between any pair of vertexes in $\hat V'$. The set of hyperarcs $\hat L \subseteq \hat A$ is called a {\it schedulable hyperarc set} if the corresponding vertex set $\hat V_{\hat L} \subseteq \hat V$ is independent. This is because the hyperarcs in $\hat L$ is conflict-free and could be scheduled simultaneously in network coding manner when $\hat V_{\hat L}$ is independent in $\hat G$. Let $\hat{\cal I}$ be the collection of all independent sets in $\hat G$, i.e., $\hat{\cal I}$ consists of all schedulable hyperarc sets of $\hat A$.
Therefore, we use $(N,\hat A,\hat G,\hat{\cal I})$ to represent a multihop wireless network with network coding in this paper.

The multihop wireless network without network coding $(N,A,G,{\cal I})$ is called the {\it original network} of $(N,\hat A,\hat G,\hat{\cal I})$. The relationship between them follows in details. Let $(i,J) \in \hat A$ be a hyperarc and $\hat v \in \hat V$ be the vertex which it corresponds to. The weights of $(i,J)$ or $\hat v$, denoted by $w(i,J)$ or $w(\hat v)$ respectively, are defined by $|J|$ or the number of sub-links. Then $A=\{(i,J): (i,J)\in \hat A,\,w(i,J)=1\}$, or $A\subseteq \hat A$ consists of all hyperarcs of $\hat A$ with weight 1. Similarly, $G=(V,E)$ is a subgraph of $\hat G=(\hat V,\hat E)$, where $V\subseteq \hat V$ consists of all vertexes of $\hat V$ with weight 1 and $E\subseteq \hat E$ consists of all edges between any pair of vertexes within $V$, and ${\cal I} \subseteq \hat{\cal I}$ is the collection of all independent sets within $G$.

Let $\hat L$ be a schedulable hyperarc set of $(N,\hat A,\hat G,\hat{\cal I})$. The set of links $L \triangleq \{ (i,j) : (i,J)\in \hat L,\, j\in J\}\subseteq A$ is called a {\it schedulable sub-link set} of $\hat L$. Let $\cal L$ be the collection of all schedulable sub-link sets of $(N,\hat A,\hat G,\hat{\cal I})$.
Arrange the links of $A$ by a fixed order and index it by $1,2,\ldots, |A|$. For $L\subseteq A$, its incidence vector is a binary vector with length $|A|$ and the $i$-th component is 1 if the $i$-th link is in $L$ and 0 otherwise for $i=1,2,\ldots,|A|$. Let $\hat P$ be the convex hull of the incidence vectors of all schedulable sub-link sets in $\cal L$, called the \emph{independence polytope} of $(N,\hat A,\hat G,\hat{\cal I})$.

A {\it fractional hyperarc schedule}, denoted by $\hat{\cal S}$, is defined by a sequence of pairs $\{({\hat L}_j,\lambda_j) \in \hat{\cal I}\times {\mathbb R}_+:1 \le j \le k\}$, where $k$ is a positive integer and $\sum _{j=1}^k \lambda_j \le 1$. The value $\sum _{j=1}^k \lambda_j$ is called the {\it length} of $\hat{\cal S}$. The \emph{link capacity function} $c_{\hat{\cal S}} \in \mathbb R_+^A$ of $\hat{\cal S}$ is defined by
\begin{eqnarray}
c_{\hat{\cal S}}(a)=\sum_{j=1}^k \lambda_j|L_j \cap \{a\}|=\sum_{j: \,a\in L_j} \lambda_j,\quad\forall a \in A,
\end{eqnarray}
where $L_j$ is the schedulable sub-link set of $\hat L_j,\;j=1,\ldots,k$.
A link demand function $d \in {\mathbb R}_+^A$  is said to be \emph{achievable} if there exists a fractional hyperarc schedule $\hat{\cal S}$ such that $c_{\hat{\cal S}}(a)=d(a)$ for any $a\in A$, and $\hat{\cal S}$ is called a fractional hyperarc schedule of $d$. For an achievable link demand function $d$, the fractional hyperarc schedule of $d$ with minimum length is called an {\it optimal fractional hyperarc schedule} of $(\hat G,d)$, and the minimum length is denoted by $\chi_f (\hat G,d)$. Clearly, the link demand function $d$ could be regarded as a vector $(d(a),a\in A)$ with length $|A|$. The {\it schedulable polytope} of $(N,\hat A,\hat G,\hat{\cal I})$ is defined by the set of achievable link demand functions. With some similarity to Proposition \ref{pro1} for the network $(N,A,G,\cal I)$ in form, we have the following results for the network $(N,\hat A, \hat G, \hat{\cal I})$, where the proofs are omitted due to the limitation of space.

\begin{Theorem}\label{th1}
For a multihop wireless network with network coding $(N,\hat A,\hat G,\hat{\cal I})$ under the protocol model, its schedulable polytope is equivalent to its independence polytope $\hat P$.
\end{Theorem}

\begin{Remark}\label{rem1}
Consider $(N,\hat A,\hat G,\hat{\cal I})$ together with its original network $(N,A,G,{\cal I})$. Let $P$ and $\hat P$ be respectively the schedulable polytopes of $(N,A,G,{\cal I})$ and $(N,\hat A,\hat G,\hat{\cal I})$. Then, it could be shown that $P \subset \hat P$. Based on Proposition \ref{pro1} and Theorem \ref{th1}, both $P$ and $\hat P$ are the convex hulls of the incidence vectors of all schedulable link sets (in the form of schedulable sub-link set for $\hat P$). The schedulable link set of $(N,A,G,{\cal I})$ is indicated by ${\cal I}$ directly, while that of $(N,\hat A,\hat G,\hat{\cal I})$ is indirectly indicated by $\hat{\cal I}$ via the mapping from hyperarc to sub-links. Any schedulable link set of $(N,A,G,{\cal I})$ is also schedulable in $(N,\hat A,\hat G,\hat{\cal I})$. But for any schedulable hyperarc set of $(N,\hat A,\hat G,\hat{\cal I})$ that includes a hyperarc with $k>1$ sub-links, its schedulable link set (in the form of schedulable sub-link set) is not schedulable in $(N,A,G,{\cal I})$.
\end{Remark}
\section{Maximum Multiflow with Network Coding}
In this section, we explain the MMF problem in wireless network coding. Consider a multihop wireless network with network coding $(N,\hat A,\hat G,\hat{\cal I})$ together with its original network $(N,A,G,{\cal I})$.
Let $\hat P$ and $P$ be respectively the schedulable polytopes of $(N,\hat A,\hat G,\hat{\cal I})$ and $(N,A,G,{\cal I})$. The polytope $\hat P$, consisting of all achievable link demand functions that can be achieved by some optimal fractional hyperarc schedule, is also called the capacity region of $(N,\hat A,\hat G,\hat{\cal I})$.

A $k$-flow $<f_1,f_2,\cdots ,f_k>$ is said to be {\it schedulable} in $(N,\hat A,\hat G,\hat{\cal I})$ if it satisfying $\sum _{i=1}^k f_i \in \hat P$. Therefore, the MMF problem in wireless network coding can be formulated by the following linear program problem,
\begin{eqnarray*}
\mbox{(LP-II)}&&\max\; \sum_{i=1}^k val(f_i), \\
\mbox{s.t.} && f_i \in \mathcal{F}_i, \;i=1,2,\ldots,k, \quad \sum _{i=1}^k f_i \in \hat P.
\end{eqnarray*}

The maximum throughput in wireless network coding is achieved by the MMF problem (LP-II). By Remark \ref{rem1}, $P \subset \hat P$, which implies that network coding could enlarge the capacity region of multihop wireless network and potentially improve the maximum throughput given by MMF problems. Fig.\ref{fig3} illustrates an example of multihop wireless network with network coding, where the original network is illustrated in Fig.\ref{fig1} and the maximum throughput is improved from $1/2$ to $2/3$ by a simple `XOR' coding operation.

Moreover, we establish the following hardness result, where $\hat {\cal N}$ is the class of multihop wireless networks with network coding discussed in this paper, and the proofs are omitted due to the limitation of space.

\begin{Theorem}\label{th2} Even restricted to the subclass of $\hat {\cal N}$, in which all nodes have uniform (and fixed) communication radii and interference radii, the positions of all nodes are available, and the coding structure is fixed, finding an optimal fractional hyperarc schedule to meet a given achievable link demand function is NP-hard. \end{Theorem}

\begin{figure}[t]
\centering{
\includegraphics[width=2.8in]{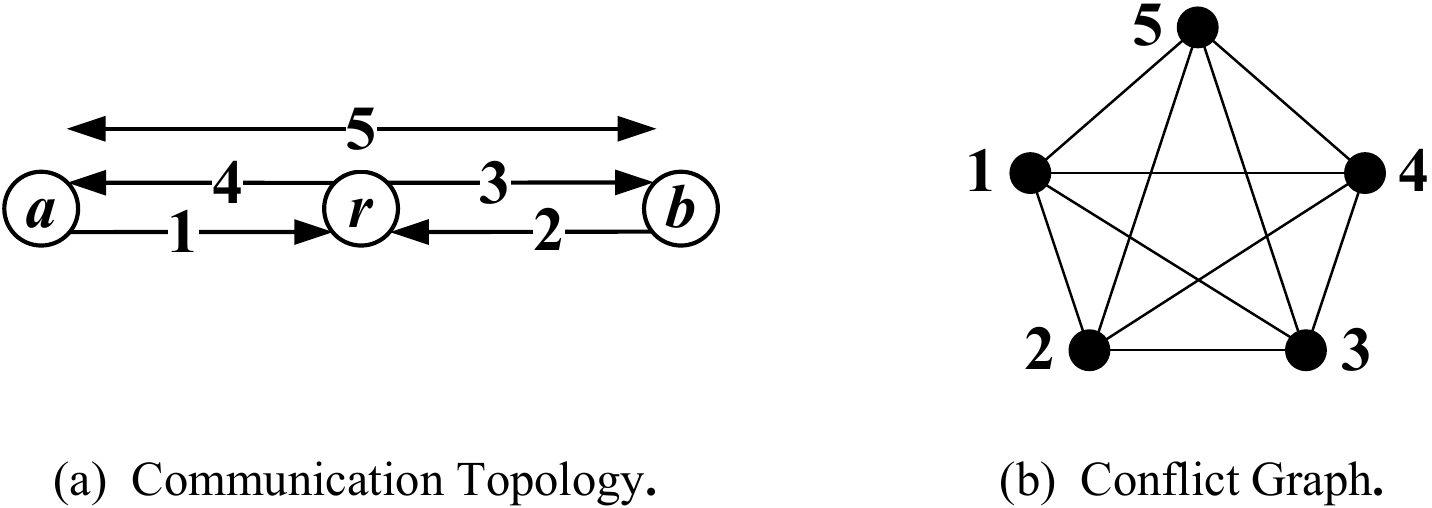}
}
\caption{An example with two commodities $\{a,b\}\{b,a\}$ which are supported by a multihop wireless network with network coding $(N,\hat A,\hat G,\hat{\cal I})$, where $N=\{a,r,b\}$, $\hat A=\{1,2,3,4,5\}$ and $\hat{\cal I}=\{\{1\},\{2\},\{3\},\{4\},\{5\}\}$. Note that the encoded transmission `5' of $\hat A$ or $(r,\{b,a\})$ is a hyperarc and its sub-link set is $\{(r,b),(r,a)\}$ or $\{3,4\}$. Then, ${\cal L}=\{\{1\},\{2\},\{3\},\{4\},\{3,4\}\}$ and the capacity region $\hat P$ is the convex hull of the following 5 incidence vectors (1000)(0100)(0010)(0001)(0011). By formulating the MMF problem, the resulting maximum throughput is $2/3$ with $val(f_1)=1/3$ and $val(f_2)=1/3$. An optimal fractional hyperarc schedule to achieve the maximum throughput is $\hat{\cal S}=\{(1,\frac{1}{3}),(2,\frac{1}{3}),(5,\frac{1}{3})\}$.}
\label{fig3}
\end{figure}

\section{Approximation Algorithm}
In this section, we give a skeleton of our algorithm, called {\it Coding-First Scheduling} (CFS), which is to find an approximate fractional hyperarc schedule in $(N,\hat A,\hat G,\hat{\cal I})$ to meet a given link demand function $d \in {\mathbb R}_+^A$.

Let $\hat G=(\hat V,\hat E)$ and $G=(V,E)$ be respectively the conflict graphs of $(N,\hat A,\hat G,\hat{\cal I})$ and its original network $(N,A,G,{\cal I})$. Let $\hat v^{(k)} \in \hat V$ denote a vertex with weight $k$, i.e., $w(\hat v)=k$, and $\hat V^{(k)}$ be the set of weight-$k$ vertexes in $\hat V$. Clearly, $\hat v^{(1)}\in G$ and $\hat V^{(1)}=V$. Let $m=|\hat A|=|\hat V|$, $n=|A|=|V|$ and $W=\max _{\hat v \in \hat V} w(\hat v)$.
An independent set $\hat I\in \hat{\cal I}$ is called a {\it maximum weighted independent set} (MWIS) of $(\hat G,w)$ if
$$w(\hat I)\triangleq \sum_{\hat v\in \hat I} w(\hat v)=\max_{\hat I'\in \hat{\cal I}}w(\hat I').$$
A vertex ordering of $\hat V$ is called a {\it coding-first ordering} of $(\hat G,w)$ if all vertexes $\hat v^{(k)} \in \hat V$ are in descending order of $k$ and in arbitrary order within $\hat V^{(k)}$, denoted by $<\hat V^{(W)},\ldots,\hat V^{(k)},\ldots,\hat V^{(2)},V>$. Let $\Omega$ be a coding-first ordering of $(\hat G,w)$.
For any $\hat U \subseteq \hat V$, a {\it coding-first MWIS} of $\hat U$ under $\Omega$ is selected as follows:
\begin{enumerate}
\item Start with an independent set $\hat I=\emptyset$ of $\hat G$;
\item Under the ordering $\Omega$, add the first vertex of $\hat U$ into $\hat I$;
\item Traverse $\hat U$ in the ordering $\Omega$. Add a new vertex into $\hat I$ if it does not have an edge in $\hat G$ to any of the vertexes added to $\hat I$ so far. At last, output $\hat I$.
\end{enumerate}
Note that the output $\hat I$ of the coding-first MWIS algorithm is an approximate MWIS.

CFS runs in iterations (see Algorithm 1). At the beginning of each iteration, it assigns the demands of links $v \in \hat U$  to hyperarcs $\hat v^{(k)} \in \hat U$ by $d(\hat v^{(k)}) \gets \min_{v \triangleright \hat v^{(k)}} d(v)$, where $v \triangleright \hat v^{(k)}$ denotes that the link $v$ is a sub-link of the hyperarc $\hat v^{(k)}$. The process of demand assignment implies that the potential demand for a hyperarc (i.e, transmitting in network coding manner) is no more than any residue demand of the sub-link it contains. CFS counts the residue demand of links $v \in \hat U$ at the end of each iteration.

\begin{algorithm}
\caption{Coding-First Scheduling (CFS)}
\label{CFS}
\begin{algorithmic}
\REQUIRE $(N,\hat A,\hat G,\hat{\cal I})$ with $\hat G=(\hat V,\hat E)$, $\Omega$ \AND $d={\mathbb R}_+^V$
\ENSURE a fractional hyperarc schedule $\Pi$
\STATE $\Pi \leftarrow \emptyset$; $\hat U \leftarrow \hat V$;
\REPEAT
\FORALL{$\hat v^{(k)} \in \hat U, k>1$}
\STATE $d(\hat v^{(k)}) \leftarrow \min_{v \triangleright \hat v^{(k)}} d(v)$;
\ENDFOR
\STATE remove $\hat v \in \hat U$ with $d(\hat v)=0$ from $\hat U$;
\STATE $\hat I \leftarrow$ coding-first MWIS of $\hat U$;
\STATE $\lambda \leftarrow \min_{\hat v \in \hat I}d(\hat v)$;
\STATE add $(\hat I,\lambda)$ into $\Pi$;
\FORALL{$\hat v \in \hat I$}
\STATE $d(v) \leftarrow \{d(v) - \lambda : v \triangleright \hat v\}$;
\ENDFOR
\UNTIL{$\hat U=\emptyset$}
\STATE Output $\Pi$;
\end{algorithmic}
\end{algorithm}

At each iteration, CFS selects a coding-first MWIS of remaining vertexes and add it into the hyperarc schedule with a demand to satisfy at least one link $v \in \hat U$. Therefore, CFS runs in at most $n$ iterations, and its running time is $O(nm)$. Theorem \ref{th5} gives a bound on the length of the output schedule.

\begin{Theorem}\label{th5}The length of the scheduling output by CFS is upper bounded by $\max _{1 \le i \le n} d(V_i)$, where $V_i$ consists of $v_i$ and all its neighbors within $V=\{v_1,\ldots,v_n\}$.\end{Theorem}

Let $\hat P$ be the schedulable polytope of $(N,\hat A,\hat G,\hat{\cal I})$, consisting of $\{d \in {\mathbb R}_+^V:\chi_f(\hat G,d)\le 1\}$. By the ordering $\Omega$, the {\it inductive schedulable polytope} $\hat P_A$ of $(N,\hat A,\hat G,\hat{\cal I})$ is defined by $\{d \in {\mathbb R}_+^V:\max_{1 \le i \le n}d(V_i) \le 1\}$. The {\it inductive schedulable number} $\alpha$* is defined by
\begin{eqnarray}
\alpha^*=\max_{1 \le i \le n,L\in \cal L}|L\cap V_i| .
\end{eqnarray}
It could be shown that $\hat P_A$ is an $\alpha^*$-approximation of $\hat P$, i.e.,
\begin{eqnarray}
\hat P_A \subseteq \hat P \subseteq \alpha^*\hat P_A,
\end{eqnarray}
and the length output by CFS is upper bounded by $\alpha^*\chi _f(\hat G,d)$. Moreover, suppose any link in $A$ conflicts with at most $\Delta$ other links in $A$, then $\alpha^* \le \Delta$, which implies the following result.

\begin{Theorem}\label{th6}For a multihop wireless network with network coding under the protocol model, where a link conflicts with at most $\Delta$ other links, the MMF problem (LP-II) has a polynomial $\Delta$-approximation algorithm.
\end{Theorem}

\section{Conclusions and Discussions}
By studying the impacts of network coding on wireless interferences, it is observed that nodes with coding operations could equivalently overcome the interference of involved transmissions and thus improve the throughput of multihop wireless networks. Based on these ideas, we propose a network model, a framework of computing the maximum throughput and a polynomial approximation algorithm for multihop wireless networks with network coding. Some discussions follow.

{\it Generalization}. The proposed framework can be generalized in many ways, e.g, it could be extended to the general settings with arbitrary link bandwidth function $b \in {\mathbb R}_+^A$ as follows. For any $f \in {\mathbb R}_+^A$, $f\oslash b\in {\mathbb R}_+^A$ is defined by $(f\oslash b)(a)=f(a)/b(a)$ for any $a\in A$. Then the MMF problem could be formulated by replacing the constraint $\sum _{i=1}^k f_i \in \hat P$ with $(\sum _{i=1}^k f_i) \oslash b \in \hat P$ in LP-II. Other generalizations could be multiple channels multiple radios, multiple transmitting power levels, other interference models, other objective functions, etc.

{\it Limitation}. The proposed framework, based on any given communication topology of wireless network coding, is unfit for networks with variable topology or coding structure. Furthermore, the proposed scheduling algorithm implies a centralized scheduling entity, which is sometimes difficult to be realized.

Our future works will focus on the generalization and the realization of the proposed framework.

\section*{Acknowledgment}
The authors would like to thank Dr. Zhi-Guo Wan for valuable discussions that help to greatly improve the paper.

\end{document}